\documentclass[superscriptaddress,longbibliography,preprint]{revtex4-1}
\usepackage{graphicx}
\usepackage{color}
\usepackage{amsmath}

\begin{document}

\title{{Using machine learning to identify factors that govern amorphization of irradiated pyrochlores}}

\author{Ghanshyam Pilania}
\affiliation{Materials Science and Technology Division, Los Alamos National Laboratory, Los Alamos, NM, USA}
\author{Karl R. Whittle}
\affiliation{School of Engineering, University of Liverpool, Liverpool, UK}
\author{Chao Jiang}
\affiliation{Fuels Modeling and Simulation Department, Idaho National Laboratory, Idaho Falls, ID, USA}
\author{Robin W. Grimes} 
\affiliation{Department of Materials, Imperial College London, London, UK SW7 2AZ}
\author{Christopher R. Stanek}
\affiliation{Materials Science and Technology Division, Los Alamos National Laboratory, Los Alamos, NM, USA}
\author{Kurt E. Sickafus}
\affiliation{Department of Materials Science and Engineering, University of Tennessee, Knoxville, TN, USA}

\author{Blas Pedro Uberuaga}
\affiliation{Materials Science and Technology Division, Los Alamos National Laboratory, Los Alamos, NM, USA}
\email{blas@lanl.gov}

\date{\today}

\begin{abstract}

Structure-property relationships is a key materials science concept that enables the design of new materials. In the case of materials for application in radiation environments, correlating radiation tolerance with fundamental structural features of a material enables materials discovery. Here, we use a machine learning model to examine the factors that govern amorphization resistance in the complex oxide pyrochlore ($A_2B_2$O$_7$). We examine the fidelity of predictions based on cation radii and electronegativities, the oxygen positional parameter, and the energetics of disordering and amorphizing the material. No one factor alone adequately predicts amorphization resistance. We find that, when multiple families of pyrochlores (with different B cations) are considered, radii and electronegativities provide the best prediction but when the machine learning model is restricted to only the $B$=Ti pyrochlores, the energetics of disordering and amorphization are optimal. This work provides new insight into the factors that govern the amorphization susceptibility and highlights the ability of machine learning approaches to generate that insight.

\end{abstract}
\maketitle

\begin{figure} 
\includegraphics[width=3.5in]{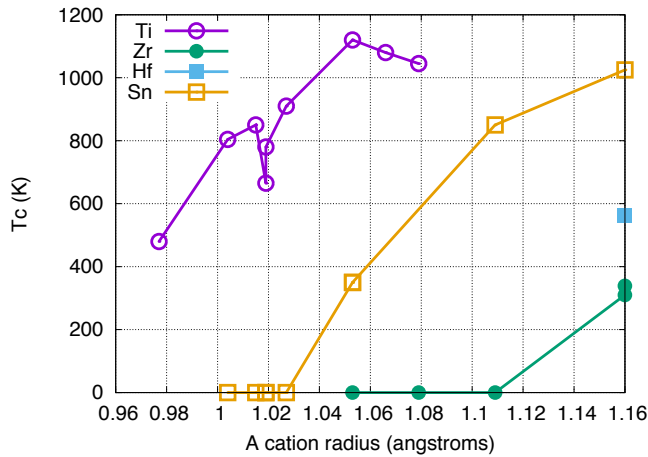}
\caption{\label{Tc-vs-rA} {
Experimentally measured values of $T_C$, ordered as a function of $A$ cation radius, for several different pyrochlores.
}}
\end{figure}

Designing materials for advanced or next-generation applications requires understanding of how properties are related to structure, thatis, identifying so-called structure-property relationships. Having such relationships guides the search for new materials with enhanced performance by identifying regions of structure and composition space that exhibit superior properties. For nuclear energy materials, a key performance metric is tolerance against radiation damage. Pyrochlores ($A_2$$B_2$O$_7$) have been extensively studied for their potential application as nuclear waste forms~\cite{Sickafus00,Begg01,Lian03a,Lian03b,Lian2006,Helean04,Sickafus07,Lumpkin07,Sattonnay13,Li12} and have been incorporated into some compositions of the SYNROC waste form~\cite{synroc}. In this context, significant effort has been directed toward understanding how the chemistry of the pyrochlore -- the nature of the $A$ and $B$ cations -- dictates the amorphization susceptibility of the compound. In particular, several experimental efforts~\cite{Ewing03,Lumpkin04,Lumpkin09,Whittle11,Lian11} have been focused on determining the critical amorphization temperature, $T_C$, the temperature at which the material recovery rate is equal to or faster than the rate of damage, as summarized in Fig.~\ref{Tc-vs-rA}. Typically, these experiments were performed in an electron microscope equipped with an ion source, such that samples were simultaneously irradiated with electrons and 1 MeV Kr ions. Though the value of $T_C$ is expected to vary depending on ion irradiation conditions~\cite{Meldrum2002}, 1 MeV Kr ion irradiation results should be comparable.

As a consequence, a number of ``features'' -- or basic structural and energetic properties -- have been identified that provide insight into the radiation response of pyrochlores. These include the radii and electronegativities of the $A$ and $B$ cations~\cite{Lumpkin04,Lumpkin07}; the $x$ parameter, which describes how the oxygen sublattice deviates from ideality~\cite{Lian03b,Lumpkin04,Lumpkin07}; the enthalpy of formation of the pyrochlore~\cite{Helean04,Helean04b}; and the energy to disorder the pyrochlore to a disordered fluorite structure~\cite{Sickafus00,Minervini00}. Further, there has been discussion on the extent of the disordered phase field in the phase diagram and its relationship to amorphization resistance~\cite{Sickafus07}. Most of these features have been only heuristically correlated with amorphization resistance or only applied to a subset of pyrochlore chemistries. We are only aware of one attempt to quantify the relationship between these types of features and a prediction of $T_C$. In that work, Lumpkin and co-workers established a relationship between $T_C$ and lattice constants, electronegativies, disordering energetics, and oxygen positional parameter~\cite{Lumpkin07}. While their model provided a significant advance in describing the structure-property relationships of pyrochlores, here we demonstrate how, through the use of machine learning, greater insight can be extracted. In particular, while they considered the disordering energy as one of their features, they used data from atomistic potentials that does not adequately describe all of the chemistries in the experiments. Further, they did not have access to data describing the amorphous state of these compounds. Finally, modern machine learning methods, applied to materials science, offer new avenues to examine the structure-property relationships in these types of systems.

Here, we use machine learning methods to demonstrate how a set of features, for a range of pyrochlore chemistries, can be used to predict $T_C$. We use both structural parameters such as cation radius and electronegativity supplemented by energetics calculated with density functional theory (DFT) to build a database of features as a function of pyrochlore chemistry. We analyze this database, building machine learning models that predict $T_C$ as a function of pyrochlore chemistry based on a systematic collection of features. We consider pyrochlore chemistries for which experimental data exists for $T_C$, which includes pyrochlores where $B$=Ti, Zr, Hf, and Sn. We find that, when considering the full range of chemistries, the two features that best predict $T_C$ are the ratio of the radii and the difference in electronegativities of the $A$ and $B$ cations. However, to predict more subtle dependencies of $T_C$ with pyrochlore chemistry characteristic of a given B chemistry, the energies to disorder and amorphize the compound provide a better prediction of $T_C$.

As compared to Ti, Hf, or Zr, Sn is a chemically very different element. It, like Ti, is multivalent, but unlike Ti, has a much stronger prevalence to adopt a charge state other than 4+. Further, as discussed below, it has a significantly higher electronegativity than the other B cations, producing a more covalent bond. This implies that Sn pyrochlores should be less amorphization resistant~\cite{Naguib1975}. However, experiments have shown Sn pyrochlores to be more amorphization resistant than other pyrochlores~\cite{Lian2006}. This all suggests that Sn pyrochlores are electronically much more complex than the other pyrochlore families, which is one reason that we use DFT to determine the energetics of disordering and amorphization, as DFT can account for the varied valence of the Sn cations. Further, the inclusion of Sn pyrochlores in this analysis, precisely because the behavior is counter-intuitive, provides a more stringent test of the methodology.

\section*{Results}
\subsection*{DFT Energetics}

\begin{figure} 
\includegraphics[width=3.5in]{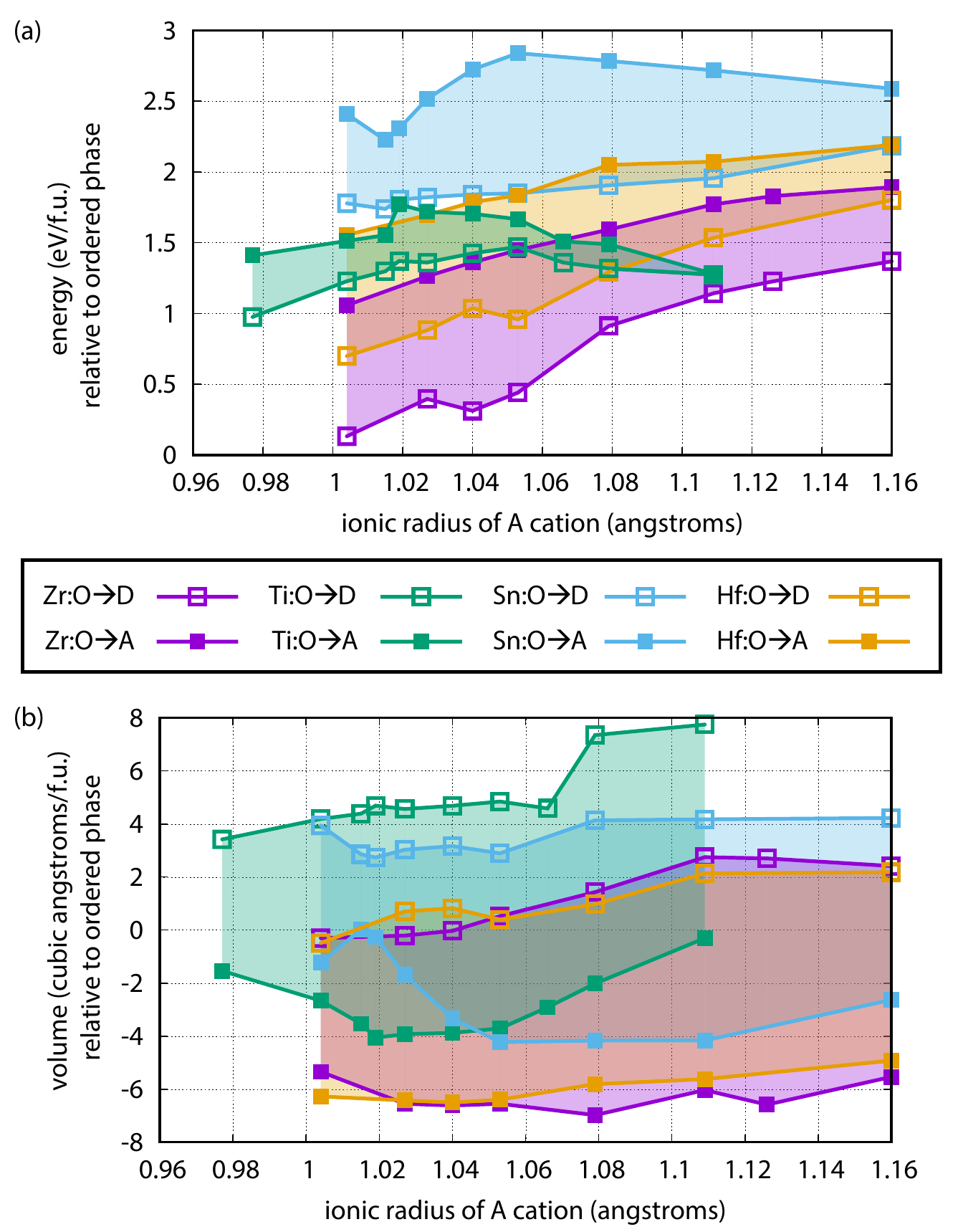}
\caption{\label{DFT-results} {
DFT results for the (a) energetics and (b) volume changes associated with an order-to-disorder (O$\rightarrow$D, open symbols) and an order-to-amorphous (O$\rightarrow$A, closed symbols) transformation for four families of pyrochlores in which $B$=Ti (green), Zr (purple), Hf (yellow), and Sn (cyan). The shaded regions highlight the differences between the disordered and amorphous structures.
}}
\end{figure}

Figure~\ref{DFT-results}a provides the energetics for disorder and amorphization of a given pyrochlore, as found using DFT, as a function of the chemistry of the pyrochlore. These are ordered by $A$ cation radius. Focusing first on the energetics to disorder, there is a general trend that as the $A$ cation radius increases, the energy associated with disordering the pyrochlore to a disordered fluorite also increases, consistent with previous results using DFT~\cite{Jiang09}. This is particularly true of the $B$=Zr, Hf and Sn families of pyrochlores. For the $B$=Ti family, there is a peak in the disorder energy for the $A$=Gd composition, again consistent with previous DFT and empirical potential calculations~\cite{Minervini00,Jiang09}.

Figure~\ref{DFT-results}a highlights the apparent contradiction between experimental observations and the notion that the disordering energy correlates with amorphization resistance. If only disordering energetics dictated the response of the pyrochlore to irradiation, then one would expect that Zr pyrochlores would generally exhibit higher amorphization resistance than Ti pyrochlores (which they do) but also that Sn pyrochlores would be less resistance to amorphization than Ti pyrochlores, which they are not. Thus, other factors must also be important. We propose that the energy of the amorphous phase is one of those factors.

The energy differences between ordered pyrochlore and an amorphous structure are also provided in Fig.~\ref{DFT-results}a. In the case of the $B$=Hf and Zr families, these are again relatively monotonic with increasing $A$ cation radius. However, the behavior of the $B$=Ti and Sn families is more complex. In particular, for the $B$=Ti family, the amorphous energy is non-monotonic with $A$ cation radius, but the peak is for a different chemistry than was the disordering energy. In the $B$=Ti family, the amorphous energy is greatest for $A$=Y and generally is high for $A$=Dy and Tb. The $B$=Sn family exhibits even more complicated behavior. There is a peak in the amorphous energy for $A$=Gd and a minimum for $A$=Ho.

Finally, the shaded regions in Fig.~\ref{DFT-results}a highlight the energy gap between the disordered and amorphous states. The variation of this gap with $A$ cation radius is very different for the different families of pyrochlores. For the $B$=Zr and Hf pyrochlores, the gap slowly but steadily decreases with $A$ cation radius. For the $B$=Ti pyrochlores, the gap first increases slightly and then decreases to essentially zero for the $A$=Nd chemistry. The gap for $B$=Sn pyrochlores first decreases, then increases, and then decreases again. Further, the gap is smallest for the $B$=Ti pyrochlores and, overall, largest for the $B$=Zr and Sn pyrochlores, at least for some A chemistries.

Figure~\ref{DFT-results}b provides the volume changes between the ordered phase and both the disordered and amorphous phases, as determined from the DFT calculations. For nearly all of the cases, a transformation from the ordered to disordered phase results in a volume expansion while the formation of the amorphous phase contracts the lattice. The exceptions are the $B$=Zr and Hf pyrochlores with small $A$ cations, which exhibit very little change in volume upon disordering. On the other hand, Nd$_2$Ti$_2$O$_7$ (assumed to be cubic here) exhibits very little change upon amorphization.

\subsection*{Correlation of Features with Amorphization Resistance}

The DFT results reveal that there are significant differences in the energetics of disorder and amorphization in pyrochlores as a function of both $A$ and $B$ chemistry. We use a machine learning approach to quantify the correlations between these energetics, as well as other features associated with pyrochlores, and the amorphization resistance, as characterized by $T_C$. The features considered here are $r_A/r_B$, the ratio of the ionic radii of the $A$ and $B$ cations; $\Delta X=X_B-X_A$, the difference in electronegativity of the $A$ and $B$ neutral metal atoms ($X_A$ and $X_B$, respectively); $x$, the oxygen positional parameter, which measures the deviation of the oxygen sublattice from an ideal (fluorite-like) simple cubic sutlattice; $E_{O\rightarrow D}$, the energy difference between the disordered and ordered phases; and $E_{D\rightarrow A}$, the energy difference between the amorphous and disordered phases. These features were chosen because (a) they have been shown to correlate to some degree in previous studies and (b) our DFT results indicate that the energetics depend strongly on the A and B chemistry of the pyrochlore, suggesting they may provide a strong descriptor of each compound. We did not consider the enthalpy of formation, proposed by other authors as a factor in radiation tolerance~\cite{Helean04,Helean04b}, as a feature because data was not available for all compounds. 

\begin{figure*} 
\includegraphics[width=7in]{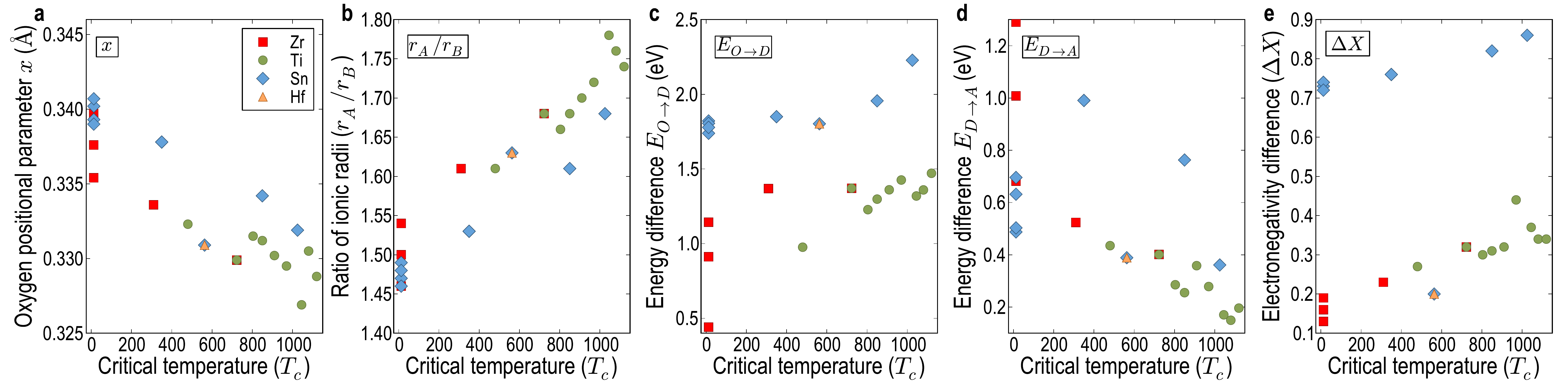}
\caption{\label{features} {
Correlations between experimental measurements of $T_C$ and 5 features describing pyrochlores. (a) $x$, the oxygen positional parameter. (b) $r_A/r_B$, the ratio of cation radii. (c) $E_{O\rightarrow D}$, the energy difference between disordered fluorite and ordered pyrochlore. (d) $E_{D\rightarrow A}$, the energy difference between an amorphous structure and disordered fluorite. (e) $\Delta X$, the difference in electronegativity between the $A$ and $B$ cation. The different symbols in the figures indicate the nature of the $B$ cation (squares=Zr, circles=Ti, diamonds=Sn, triangles=Hf).
}}
\end{figure*}

However, before we examine the results of the machine learning model, it is instructive to examine how the selected features correlate with $T_C$. Figure~\ref{features} provides simple plots of each feature against $T_C$. The values for $T_C$, summarized in Table S1, are taken from Refs.~\cite{Ewing03,Lumpkin04,Whittle11,Lian11}. Figure~\ref{features} reveals that while there are rough correlations between $T_C$ and some of the features, there is not one feature that provides a quantitative capability of predicting $T_C$. For example, overall, $r_A/r_B$ correlates well with $T_C$ over a wide range of $B$ chemistries; however, it does not capture subtleties associated with variations in $T_C$ with a given family of pyrochlores. $\Delta X$, on the other hand, discriminates between pyrochlores with $B$=Sn and the other families but does not correlate directly with $T_C$. Similarly, $x$ shows an overall correlation with $T_C$ but again the details are lost. $E_{O\rightarrow D}$, on the other hand, seems to correlate reasonably well for pyrochlores within a given family but does not describe variations of $T_C$ between families. Finally, $E_{O\rightarrow A}$, similar to $x$ and $\Delta X$, seems to generally correlate separately for $B$=Sn pyrochlores and the other families of pyrochlores. Thus, while there are rough trends indicating some insight from each of these features, there is certainly not enough of a correlation in any case for a quantitative prediction. However, this suggests, as noted by other authors~\cite{Lumpkin07}, that combinations of these features may provide predictive capability. Hence, we use a machine learning approach to quantify this.

\subsection*{Results of the Machine Learning Model}

We use a machine learning (ML) approach to quantify the correlations between the five features described in the previous section and $T_C$. More specifically, we employed kernel ridge regression (KRR)\cite{KRR-Witten,KRR-Muller,KRR-Hofmann}---an algorithm that works on the principle of similarity and is capable of extracting  complex non-linear relationships from data in an efficient manner---with a Gaussian kernel to learn and quantify trends exhibited by $T_C$ in the feature space discussed above. A randomly selected 90\%/10\% training/test split of the available data was used for statistical learning and testing the performance of the trained model on previously unseen data. A leave-one-out cross validation is used to determine the model hyper-parameters to avoid any overfitting of the training data that may lead to poor generalizability. The trained model can subsequently be used to make an interpolative prediction of $T_C$ for a new material ($i.e.$, not used in the model training) $i$ by comparing its distance in feature space $d_i$ (suitably defined by a distance measure; in our case the Euclidean norm was used) with those of a set of reference training cases for which the $T_C$ values are known. Further details of our KRR-based ML models are provided in the Methods section.

\begin{figure*} 
\includegraphics[width=7in]{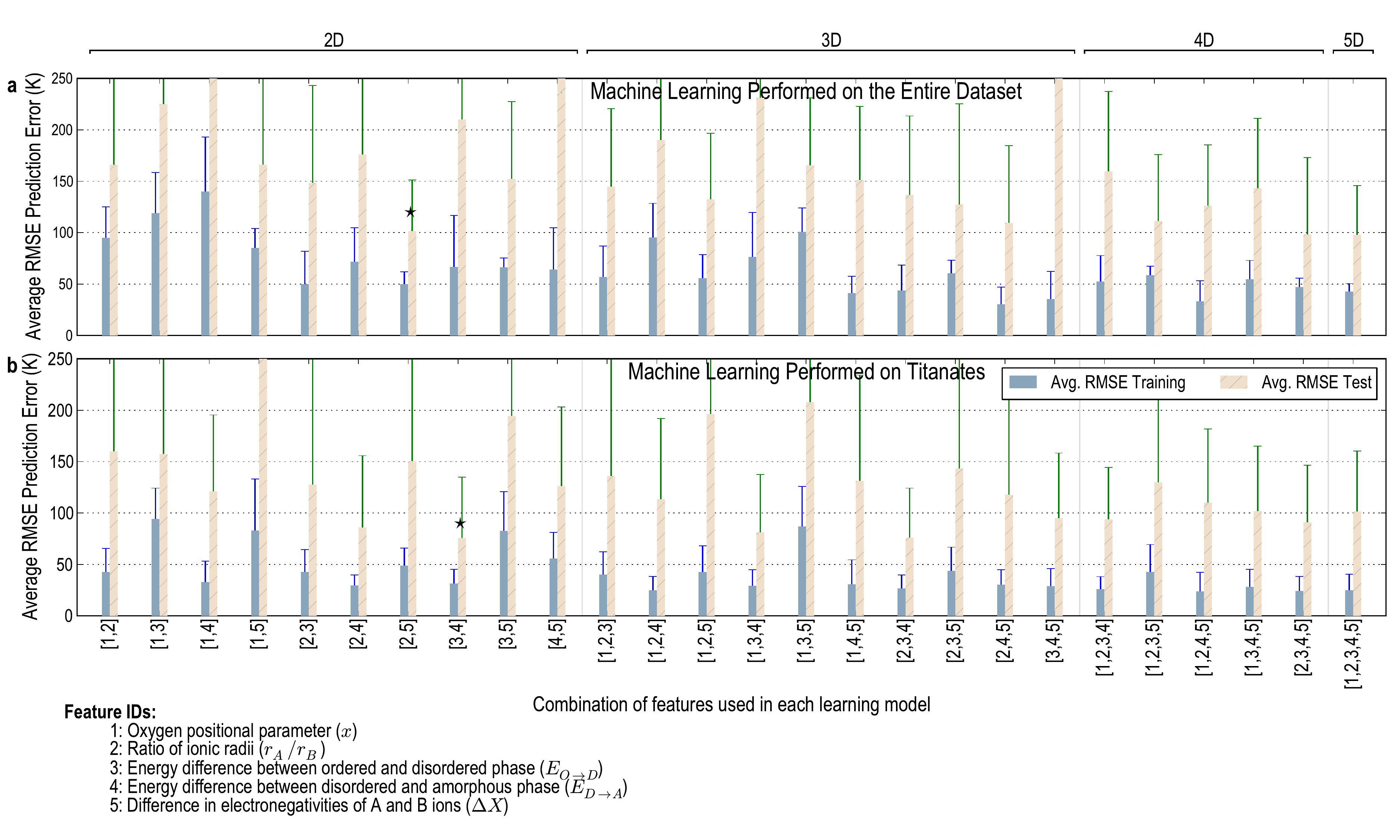}
\caption{\label{machine-learning} {
Results from the machine learning model. (a) Results when applied to the entire set of pyrochlores and (b) results when applied only to the titanate family. The solid bars indicate the average RMS error for the training data while the hashed bars indicate the error for the test data. The indices on the abscissa indicate the features used in that particular model. Models including 2, 3, 4 or all 5 features were considered. The best 2D feature set is indicated with the star for both cases. Error bars represent the standard deviations for the rms error in predict $T_C$, computed over the 100 different training/test set splits.
}}
\end{figure*}

Next, within the KRR ML model, we aim to identify the best feature combination that exhibits highest prediction performance, quantified by its ability to accurately predict $T_C$ of the test set compounds. We do this in a comprehensive manner by building KRR ML models using all possible combinations of $\Omega$ features with $\Omega$ $\in$ [2,5]. Performance of each of these models was evaluated separately on the entire data set as well as on a reduced set that only included the Ti pyrochlores. The root mean square ($rms$) errors for the $T_C$ predictions on training and test sets for various models is presented in Fig.~\ref{machine-learning}. In order to account for model prediction variability associated with randomly selected training/test splits, Fig.~\ref{machine-learning} reports the $rms$ errors averaged over 100 different randomly selected training/test splits for each of the models. The 2D models that lead to the lowest $rms$ errors on the test set data have been marked with a `$\star$' in Fig.~\ref{machine-learning}a (when taking the entire data) and Fig.~\ref{machine-learning}b (for the Ti pyrochlores). It is interesting to note that, for both cases, going beyond the best performing 2D models does not lead to a significant improvement in the model prediction performance. For instance, while the best binary feature pair ($r_A/r_B$, $\Delta X$) leads to a test set $rms$ error of 101.2 K in $T_C$, the ML models built on the best 4D and 5D (taking all 5 features considered) feature vectors only result in nominal improvements leading to $rms$ errors of 97.8K and 98K, respectively. Since, as a general rule, higher model complexity often leads to poor generalizability, in case of a comparable prediction performance, a simpler model ($i.e.$, built on a lower dimensional feature set) should always be preferred over a more complex one. Therefore, henceforth we focus our attention on the the best performing 2D models. 

The superior performance exhibited by the ($r_A/r_B$, $\Delta X$) feature pair is not entirely unexpected and can be understood by looking at Fig.~\ref{features}b and e. As alluded to previously, while $r_A/r_B$ helps capture the overall $T_C$ trends among different chemistries, $\Delta X$ allows for an effective separation between different chemistries (especially, between the Sn-based compounds and rest of the dataset), while still capturing relative $T_C$ trends between these subgroups. The best performing feature pair for the titanate pyrochlores dataset, however, is constituted by $E_{O\rightarrow D}$ and $E_{D\rightarrow A}$. While the ($r_A/r_B$, $\Delta X$) feature pair performs much poorer on this subset than the overall dataset, the performance of ($r_A/r_B$, $E_{O\rightarrow D}$) feature pair is also found comparable to that of the best 2D feature pair. 

\begin{figure} 
\includegraphics[width=3.5in]{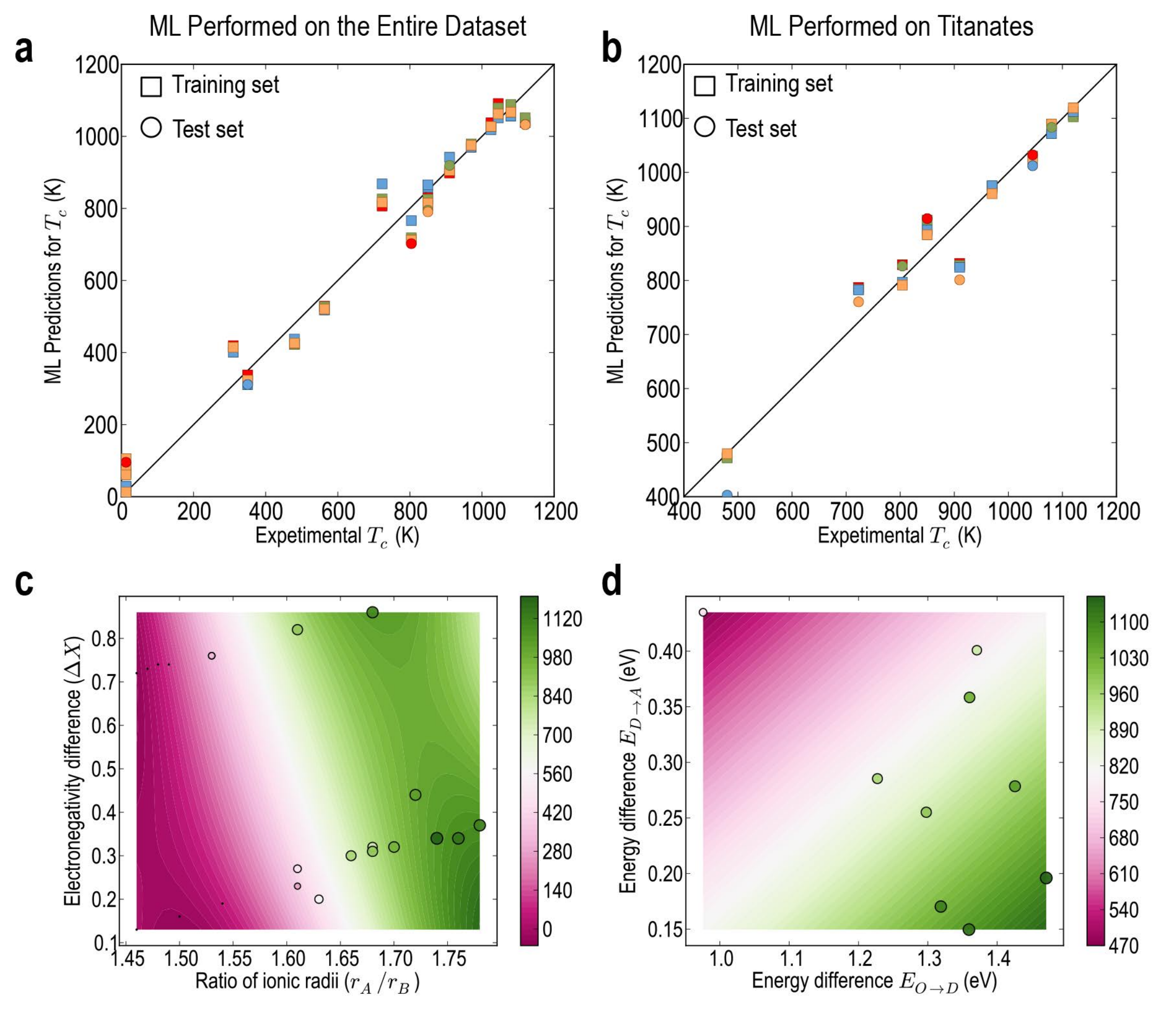}
\caption{\label{contour-plots} {
(a-b) Parity plots of the machine learning results for (a) the entire set of pyrochlores and (b) the titanate family. The squares represent training data while the circles are the test set. The different colors represent different runs with different training/test set splits.
(c-d) Results from the machine learning model for (c) all of the pyrochlores considered and (d) just the titanate famliy. The size and color of the circles indicate the experimental $T_C$ while the position of the circles indicates the predicted $T_C$. The contours indicate predictions of $T_C$ for other values of the feature pairs.
}}
\end{figure}

While Fig.~\ref{machine-learning} captures the average performance and variability (taken over 100 different runs) for our best performing 2D models (marked with a $\star$), in Fig.~\ref{contour-plots}a-b we present parity plots comparing the experimental $T_C$ with the ML predictions using the best 2D descriptors found for the entire dataset (Fig.~\ref{contour-plots}a) and the titanates (Fig.~\ref{contour-plots}b), respectively. In each case $\sim$90$\%$ of the dataset was used for training (plotted as squares) with the remaining for testing the model performance (plotted as circles). In each case, we used four different ML runs randomly selecting training and test set splits (depicted by different colors). It can be see from the parity plots that our ML models can reasonably predict (within the error bars established in Fig.~\ref{machine-learning}), $T_C$ over the entire dataset. A couple of conclusions can be drawn from these plots. First, visually it can be seen that the model prediction performance is comparable for the training and test sets, indicating that there is no overfitting (a problem when a ML model performs very well on a training set but exhibits a poor performance on a test set). Second, despite their simplicity (given that we are only using a two-dimensional feature in each case), the ML models exhibit good predictive power and stability (predictions do not change drastically over different training/test splits). This highlights the robustness of the model.

To gain a deeper insight into the KRR model's prediction performance, we next construct contour plots for each of the two best performing 2D feature pairs discussed above. In each case, we start with a fine 2D grid in the feature space constituted by the primary features identified above, while still confining ourselves within the boundaries of the original feature space used to train the KRR models. Each point on this grid then, in principle, represents a point in the feature space, which can be used as an input for the respective trained KRR models to make predictions. That is, we map out the predicted value of $T_C$ as a function of the two features over a range of values for each of the two features. Since the ML models are interpolative, one can readily use a fine grid in the 2D feature space to visualize trends in $T_C$ versus the feature values and make predictions of $T_C$ for new chemistries.

Figure~\ref{contour-plots}c shows the best two-feature descriptor for the entire set of pyrochlores considered. Again, in this case, the two features that best correlate with $T_C$ are $r_A/r_B$ and $\Delta X$. This combination of features is able to distinguish the different $T_C$ behavior exhibited by the $B$=Sn pyrochlores and the other families of pyrochlores, by virtue of the properties of $\Delta X$. 

However, as discussed above, this combination of features has an effective uncertainty of $\sim 100$ K, indicating that it cannot describe the fine features exhibited by the $B$=Ti family of pyrochlores. For example, $T_C$ is not monotonic with $A$ cation radius (see Fig.~\ref{Tc-vs-rA}). As discussed, limiting the model to just the $B$=Ti pyrochlores results in a different optimal two-feature set, namely $E_{O\rightarrow D}$ and $E_{O\rightarrow A}$, as shown in Fig.~\ref{machine-learning}b. In particular, as shown in Fig.~\ref{contour-plots}d, this set of features can describe the subtle behavior in which the $A$=Gd compound has the highest value of $T_C$, correlating with the fact that it has the highest value of $E_{O\rightarrow D}$, while the $A$=Y compound, which has values of $E_{O\rightarrow D}$ similar to the neighboring compounds, exhibits an anomalously low value of $T_C$. This is a consequence of its rather high value of $E_{D\rightarrow A}$, a consequence of the fact that Y is not a rare earth and thus the bonding associated with it is subtly different to the other elements around it. 

\section*{Discussion and Conclusions}

Combining experimental results for $T_C$ for various pyrochlore compounds, DFT calculations of the energetics of disordering and amorphization, and a machine learning model, we conclude that (a) basic ionic properties such as $r_A/r_B$ and $\Delta X$ have the qualitative capability of predicting trends in $T_C$ over a wide-range of pyrochlore compounds but that (b) more quantitative predictions that capture the subtleties associated with variations in $A$ cation chemistry require knowledge of the disordering and amorphization energetics. This generalizes the previous understanding in which rough correlations between amorphization resistance and, for example, disordering energetics were hypothesized based on a few observations. 

However, what is clear from the machine learning analysis is that, even with the input of DFT energetics, the predictive capabilities are still limited. Even when limited to the $B$=Ti family of pyrochlores, the model results in predictive uncertainty of 75 K. This is a consequence of many factors, including the limited amount of experimental data, the uncertainties in  what experimental data that there is, and uncertainties in the DFT calculations. To determine an even better predictive model, more experimental data is required. In particular, values of $T_C$ for other families of pyrochlores would enhance the strength of the model. For example, without the $B$=Sn pyrochlores, the importance of $\Delta X$ would likely not have been revealed. Importantly, given the small data set, domain knowledge -- experience with the behavior of this system -- was important in narrowing down a set of likely relevant features.

While the feature set of $\Delta X$ and $r_A/r_B$ have the best predictive capability for distinguishing between the various families of pyrochlores, the reason why Sn pyrochlores are radiation tolerant while exhibiting such high disordering energies is found in examining the amorphization energetics. The gap between the disordering and amorphization energies for the Sn pyrochlores is typically quite large and even if, during the course of irradiation, enough energy is deposited into the lattice such that the structure becomes disordered, it is not enough to amorphize the material. The gap betwen the disordering and amorphization energies is much larger in the Sn pyrochlores than it is in the Ti family and, for some A cations, larger than for the Hf and Zr families as well. Thus, the origin of the radiation tolerance of some of the Sn pyrochlores comes from the fact that they are extremely difficult to amorphize.

The insights gained by the machine learning model apply specifically to pyrochlores and, because of the interpolative nature of these models, to the families of pyrochlores considered here. That said, the features identified as being best able to predict $T_C$ can be justified physically and thus may be applicable to other classes of complex oxides, such as $\delta$-phase~\cite{Stanek2009}, that have fluorite as the parent structure. However, other classes of complex oxides, such as spinel, which have fundamentally different crystal structures may have different dependencies on these features, or require new features to predict behavior. In particular, structural vacancies on the cation sublattice in spinel can facilitate recovery of damage in a way that is not possible in pyrochlore~\cite{Uberuaga2015}. Further, other factors, such as short-range order, which is known to occur in complex oxides~\cite{Jiang12,Shamblin2016}, may also play a role. However, we suspect that treating the disordered state as truly random captures much of the behavior of these materials, given the ability of the disordered fluorite structure to predict order-disorder temperatures in these systems~\cite{Jiang09,Li2015}.  

In this work, we have used $T_C$ as a metric for relative amorphization resistance. In reality, the value of $T_C$ encompasses not only thermodynamic properties such as disordering and amorphization energetics, but also kinetic processes of defect annihilation and defect production. Thus, actually predicting $T_C$ from fundamental defect behavior would be a daunting task. However, it does provide a metric to compare the susceptibility of amorphization that has been measured for a range of pyrochlore chemistries. 

Finally, this work highlights the utility of machine learning approaches in materials science. In this case, the ML model elucidates those features which provide predictive capability, providing insight into those factors which dictate amorphization resistance in pyrochlores. The model also shows that sets of two features result in optimal predictions; higher-order feature sets do not add significant value. The fact that different combinations of features provide are optimal for predictions for the entire set of pyrochlores ($r_A/r_B$ and $\Delta X$) versus the Ti family ($E_{O\rightarrow D}$ and $E_{O\rightarrow A}$) reinforces the point that the best set of features depends on the level of detail (here, the error in the predicted $T_C$) required in the prediction.


\section*{Methods}

\subsection*{Density Functional Theory}

Density functional theory (DFT) calculations were performed using the all-electron projector augmented wave method~\cite{PAW} within the local density approximation (PBE) with the VASP code~\cite{VASP}. A plane-wave cutoff of 400 eV and dense $k$-point meshes were used to ensure convergence. The lattice parameters and all atomic positions were allowed to relax, though the cells were constrained to be cubic. The disordered fluorite structure was modeled using the special quasirandom structures (SQS) approach~\cite{SQS}. The SQS structures were generated as described in Ref.~\cite{Jiang09}. The amorphous structures were created by performing {\em ab initio} molecular dynamics at a very high temperature and then quenching the structures to 0 K.   For the $B$=Zr and Hf families, there is a deviation from true monotonic behavior at $A$=Tb, in contrast with previous DFT calculations~\cite{Jiang09} that used the same methodology (pseudopotentials, functional, k-point mesh, and energy cutoff). We assume that the differences from previously published results are due to changes in different versions of VASP.

\subsection*{Machine Learning Model}
We used Kernel ridge regression (KRR) with a Gaussian kernel for machine learning. KRR is a similarity-based learning algorithm, where the ML estimate of a target property (in our case the critical temperature $T_C$) of a new system $j$, is estimated by a sum of weighted kernel functions ($i.e.$, Gaussians) over the entire training set, as

\begin{equation}\label{eq:KRR}
T_{Cj}^{ML} = \sum \limits_{i=1}^N w_i \exp \left( -\frac{1}{2\sigma^2}|\textbf{d}^{ij}|^2 \right).
\end{equation}
where $i$ runs over the systems in the training dataset, and $|\textbf{d}^{ij}|^2 = ||\textbf{d}_i - \textbf{d}_j ||_2^2$, the squared Euclidean distance between the feature vectors $\textbf{d}_i$ and $\textbf{d}_j$. The coefficients $w_i$s are obtained from the training (or learning) process built on minimizing the expression $\sum \limits_{i=1}^N   \bigg( T_{Ci}^{ML} - T_{Ci}^{Exp} \bigg)^2  + \lambda   \sum \limits_{i=1}^N   w_i^2$, with $T_{Ci}^{ML}$ being the ML estimated critical temperature, $T_{Ci}^{Exp}$ the corresponding experimental value, and the model hyper-parameters $\sigma$ and $\lambda$ are optimized within a internal cross-validation loop on a fine logarithmic grid. The explicit solution to this minimization problem is $\boldsymbol{\alpha} = (\textbf{K}+\boldsymbol{\lambda} \textbf{I} )^{-1} \textbf{P}^{DFT}$, where $\textbf{I}$ is the identity matrix, and $K_{ij} = \exp \left( -\frac{1}{2\sigma^2}|\textbf{d}^{ij}|^2 \right)$ is the kernel matrix elements of all materials in the training set. The parameters $\lambda$ and $\sigma$ are determined in an inner loop of fivefold cross validation using a logarithmically scaled fine grid.

\section*{Acknowledgments}

This work was supported by the U.S. Department of Energy, Office of Science, Basic Energy Sciences, Materials Sciences and Engineering Division.
Los Alamos National Laboratory, an affirmative action equal opportunity employer, is operated by Los Alamos National Security, LLC, for the National Nuclear Security Administration of the U.S. DOE under contract DE-AC52-06NA25396.

\section*{Author contributions}

G.P.\ developed the machine learning model and applied it to the pyrochlores. K.R.W.\ compiled the experimental data. C.J.\ constructed the DFT simulation cells. R.W.G.\ and K.E.S.\ performed the original studies that proposed the concept linking radiation tolerance to disordering. C.R.S.\ helped devise the study. B.P.U.\ wrote the main manuscript text and performed the DFT calculations. All authors reviewed the manuscript.

\section*{Additional information}

Competing financial interests: The authors declare no competing financial interests.

\bibliography{pyrochlore-amorphization}

 \end{document}